# Self-started stable pulsing operation of random fiber laser

*Jiangming Xu[1,2], Jun Ye[1], Hanwei Zhang[1,2], Wei Liu[1], Jian Wu[1,2], Hu Xiao[1,2], Jinyong Leng[1,2], Pu Zhou[1,2]\**

[1]College of Optoelectronic Science and Engineering, National University of Defense Technology, Changsha 410073, China
[2]Hunan Provincial Collaborative Innovation Center of High Power Fiber Laser, Changsha 410073, China
\*Corresponding author: zhoupu203@163.com

**ABSTRACT**

*Unlike traditional fiber laser with defined resonant cavity, random fiber laser (RFL), whose operation is based on distributed gain and feedback via Rayleigh scattering and stimulated Raman scattering in long passive fiber, has fundamental scientific challenges in pulsing operation for its remarkable cavity-free feature. Here, we propose and experimentally realize the passively spatiotemporal gain modulation induced self-started stable pulsing operation of counter-pumped RFL. Thanks to the good temporal stability of employed pumping amplified spontaneous emission source and the superiority of this pulse generation scheme, stable and regular pulse train can be obtained. Furthermore, the pump hysteresis and bistability phenomena with the generation of high order Stokes light is presented and the dynamics of pulsing operation is discussed. This work extends our comprehension of temporal property of RFL and provides an effective novel avenue for the exploration of pulsed RFL with structural simplicity, low cost and stable output.*

## 1. INTRODUCTION

Random fiber laser (RFL), whose operation is based on the extremely weak Rayleigh scattering provided random distributed feedback (RDFB) and Raman gain in a section of passive fiber, has attracted more and more attentions in the past decades for its special features of cavity-free, mode-free, and structural simplicity and application potentials in telecommunication and distributed sensing [1-3].

In recent years, intensive investigations of RFLs have been reported on performance scaling and application exploring. However, most of the previously presented RFLs operate in continuous-wave mode with stable output. Although mode locking and time taming have been realized in random lasers formed by nanoscale particles via mode-selective pumping [4] and active control of spatial pump profile [5], pulsing operation of RFLs based on hundreds meters and even kilometers passive fiber have fundamental scientific challenges for their remarkable cavity-free feature, which is different from traditional fiber laser with defined resonant cavity and mode-locking can be obtained [6, 7]. Recently, pulsed RFLs have been demonstrated via actively modulation methods such as internal modulation utilizing electro-optical modulator [6], Q-switching with the aid of acousto-optic modulator [7], polarization modulation employing polarization switch and arbitrary waveform generator [8]. Simultaneously, passively methods, for instance, Q-switching assisted by stimulated Brillouin scattering (SBS) [9] and saturable absorption of monolayer grapheme [10], are also important schemes for the demonstration of pulsed RFLs. Furthermore, Zhang et al. [11] in our group reported the passively self-oscillation with obvious amplitude instability and continuous-wave pedestal in high order RFL, which was attributed to the depletion of the pump. It is worth to be noted that the obtaining of stable pulse from RFL based on passively method is difficult for the stochastic nature of RDFB and SBS [12].

In this manuscript, we propose and experimentally realize a self-started stable-pulsed RFL via passively spatiotemporal gain modulation. To the best of our knowledge, this is the first demonstration of stable pulsing operation of RFL by passively modulation method.

## 2. EXPERIMENTAL SETUP

The experimental setup of the counter pumped RFL is plotted in Fig. 1. Half-opened cavity is employed to decrease the threshold of Stokes light generation, which is composed by a fiber coupler with coupling ratio of 50:50 at 1080 nm and a piece of 3 km long passive fiber. By splicing the output ports of the coupler together, fiber loop mirror (FLM) can be obtained. The passive fiber spliced after the FLM provides random distributed feedback and Raman gain for the random lasing. So the feedback of this random laser is provided both by highly reflective FLM and the random distributed

Rayleigh scattering in the passive fiber. The pump source we utilized is a broadband amplified spontaneous emission (ASE) light centered at 1079.2 nm with a full width at half maximum (FWHM) linewidth of about 17 nm. The pump light is injected into the passive fiber via the first port of fiber circulator (Cir). And the random laser light is coupled out from the cavity via the third port of the circulator.

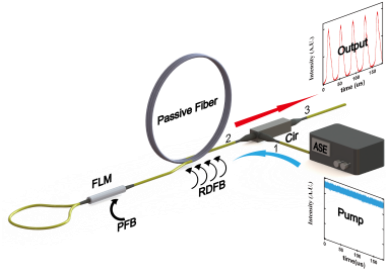

Fig. 1. Experimental setup of the counter pumped RFL. FLM: fiber loop mirror, PFB: point feedback, RDFB: random distributed feedback, Cir: circulator, ASE: amplified spontaneous emission.

## 3. RESULTS AND DISCUSSIONS

Figure 2 illustrates the output characteristics of this counter pumped RFL. Below the pump level of 6.51 W, unstable random lasing can be measured. As shown in Fig. 2(b), several high order Stokes light and lots of spikes can be observed at 5.36 W pump light injected, which is typical for the multi-cascaded Rayleigh scattering (RS)- SBS generation [2]. With the enhancement of pump power to 6.51 W, the high order Stokes light vanishes and the output spectrum becomes smooth. Furthermore, regular pulse train can be obtained at this power level, as plotted in Fig. 2(c). The period and duration of the out pulse is 37.42 μs and 10.52 μs, respectively. The period is about 1.27 times higher than the round trip time 2nL/c (n is the refraction index of fiber, L is the length of the cavity, and c is the speed of light in vacuum) in the passive fiber. This may be induced by the increment of effective length [13] of the passive fiber as the Stokes light can experience optical gain and loss simultaneously in the passive fiber. The average power of output 1st order Stokes light is measured to be 2.13 W with the aid of power meter. The corresponding peak power and pulse energy are evaluated to be about 7.06 W and 79.7 μJ via the calculation of output temporal signals.

As to the dynamics of pulsing operation, we can imagine following processes in this counter-pumped RFL. The output of Stokes light need sufficient Raman gain which is provided by the pump light distributed in the half-opened cavity. When the pump source with enough power level is on, the pump light, which experiences a round trip in the passive fiber and exports from the 3rd port of fiber circulator, would be depleted because most of its energy will transform to the 1st order Stokes light. Then the pump power distributed in the half-opened cavity at a certain moment cannot provide sufficient Raman gain for the Stokes light, and the RFL could be off. When the RFL is off, the pump light distributed in the cavity would not be depleted so much and can offer enough Raman gain for Stokes light. Then, a pulse of Stokes light is formed. At last, a kinetic balance between the pump power and Stokes power distributed in the cavity can be achieved and stable spatial-temporal modulation of pump and Stokes light can be generated. Then stable pulsing operation can be obtained. In summary, the stable pulsing operation of this counter-pumped RFL is caused by the passively spatial-temporal modulation of pump light induced self-modulation of Raman gain in the laser despite of the employing of continuous-wave pump source. The spatial-temporal distributions of pump and Stokes light will be calculated theoretically and presented at the conference.

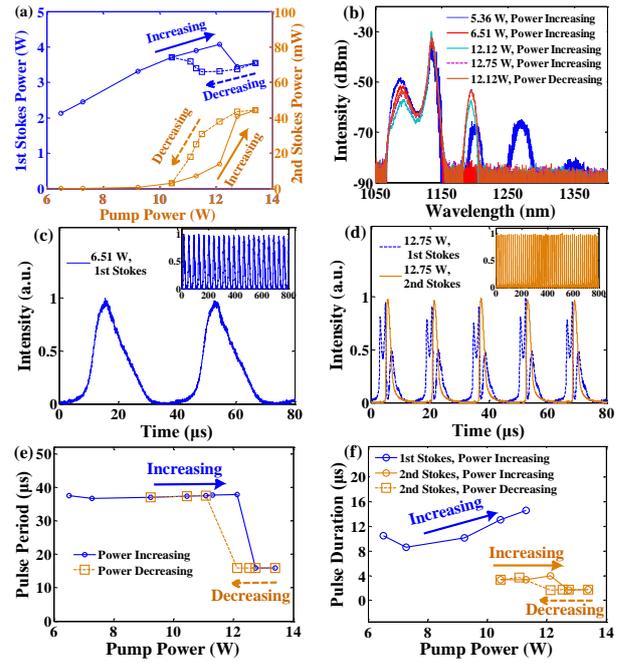

Fig. 2. Output characteristics of the counter pumped RFL. (a) The corresponding power of 1st and 2nd order Stokes light as functions of pump power. (b) Output spectra at different pump levels. (c) and (d) Temporal signals of output Stokes light at 6.51 W and 12.75 W pump levels. (e) and (f) The evolution of period and duration as functions of pump power.

With the scaling of pump power from 6.51 W to 12.12 W, tiny changing of pulse period and duration can be measured, as shown in Fig. 2(e) and Fig. 2(f). This may be attributed to the alteration of optical gain in the passive fiber with the enhancement of pump level as the output pulse is associated with the pump depletion and Raman gain in the passive fiber. At pump power of 12.12 W, the measured average power and evaluated peak power of output 1st order Stokes light is 4.07 W and 12.05 W, respectively. Furthermore, the optical peak corresponding to the 2nd order Stokes light can be observed with intensity 27.14 dB lower than the optical peak of 1st order Stokes light at this pump level.

The threshold of 2nd order Stokes light of this RFL is about 12.75 W. At this pump level, dramatically increasing of average power of 2nd order Stokes light and decreasing of average power of 1st order Stokes light can be measured, as presented in Fig. 2(a). Additionally, the difference of intensity of 1st and 2nd order Stokes light decreases to 20.89 dB.

Simultaneously, the pulse period of the 1st and 2nd order Stokes light both decrease to 15.89 μs, which is similar to the transmit time (nL/c, 14.72 μs) in the passive fiber. The changing of pulse period may also be induced by the alteration of optical gain and loss in the passive fiber, and agrees well with previously reported results in Ref. 11. Unlikely, the lasing tendency of 1st and 2nd order Stokes light from this RFL cannot be concluded in simply as anti-phase motion as described in the previous manuscript. Despite of the sharing of the same pulse period, the pulse shape of 1st and 2nd order Stokes light are not completely identical, as displayed in Fig. 2(d). For the experiencing of pulse breaking, which may be induced for the limitation of peak power in the long passive fiber, the pulse duration of 1st order Stokes light cannot be obtained exactly. And the pulse duration for 2nd order Stokes light is measured to be about 1.66 μs. The corresponding peak power/ pulse energy of 1st and 2nd order Stokes light are calculated to be about 19.69 W/ 54.66 μJ and 0.33 W/ 0.65μJ with the calculation of temporal signals, respectively.

Worthy of mention here is that the pulse period always jump at a relatively high pump power level (12.75 W, threshold of 2nd order Stokes light) in the power scaling process, and the transformation of pulse duration, output power and spectrum can be measured simultaneously. In the pump power decreasing process, the operation state could be maintained to a relatively low pump level (11.09 W) after the drastically reducing of pulse period. This phenomenon is known as pump hysteresis and bistability, which has been widely reported in mode-locked lasers [14, 15]. The pump hysteresis and bistability characteristics of this pulsed RFL may be induced by the refractivity modulation of the optical intensity [14] or the gain saturation [15], and indicate that the operation state of this laser can be influenced by not only the pump level, but also the initial state.

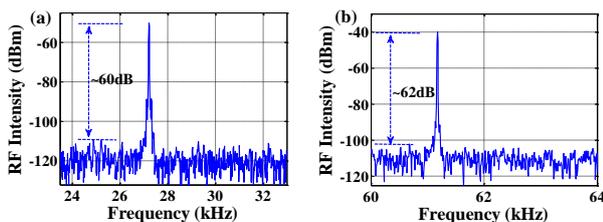

Fig. 3. RF spectra of (a) 1st order Stokes light and (c) 2nd order Stokes light around fundamental repetition frequency with a resolution of 20 Hz.

The corresponding radio-frequency (RF) spectra of the 1st and 2nd order Stokes light measured at the fundamental repetition frequency at pump power of 6.51 W and 12.75 W (threshold of 1st and 2nd order Stokes light) are presented in Fig. 3. The RF traces have contrasts of about 60 dB and 62 dB against the noise floor, respectively, which indicate the good stability of the output pulse.

## 4. CONCLUSION

In summary, we propose and realize a stable pulsed RFL via passively spatiotemporal gain modulation. Additionally, pump hysteresis and bistability can be observed with the overcoming of threshold of 2nd order Stokes light. Further pulse energy scaling and dynamical simulations of pump hysteresis and bistability are on the going. The investigations can extend the comprehension of temporal property of RFL and provide a new route for the demonstration of a pulsed RFL with structural simplicity, low cost and stable output.

## 5. REFERENCES

[1] D. S. Wiersma, "The physics and applications of random lasers," *Nat. Phys.*, vol. 4, no. 5, pp. 359–367 (2008).

[2] S. K. Turitsyn, S. A. Babin, A. E. El-Taher, P. Harper, D. V. Churkin, S. I. Kablukov, J. D. Ania-Castañón, V. Karalekas, and E. V. Podivilov, "Random distributed feedback fibre laser," *Nat. Photon.*, vol. 4, no. 4, pp. 231–235 (2010).

[3] X. Du, H. Zhang, H. Xiao, P. Ma, X. Wang, P. Zhou, and Z. Liu, "High-power random distributed feedback fiber laser: From science to application," *Ann. Der Phys.*, vol. 528, no. 9-10, pp. 649-662 (2016).

[4] M. Leonetti, C. Conti, and C. Lopez, "The mode-locking transition of random lasers," *Nature Photon.*, vol. 5, no. 10, pp. 615-617 (2011).

[5] N. Bachelard, J. Andreasen, S. Gigan, and P. Sebbah, "Taming Random Lasers through Active Spatial Control of the Pump," *Phys. Review Lett.*, vol. 109, pp. 033903 (2012)

[6] M. Bravo, M. Fernandez-Vallejo, and M. Lopez-Amo, "Internal modulation of a random fiber laser," *Opt. Lett.*, vol. 38, no. 9, pp. 1542-1544 (2013).

[7] J. Xu, J. Ye, H. Xiao, J. Leng, J. Wu, H. Zhang, and P. Zhou, "Narrow-linewidth Q-switched random distributed feedback fiber laser," *Opt. Express*, vol. 24, no. 17, pp. 19203–19210 (2016).

[8] H. Wu, Z. Wang, Q. He, M. Fan, Y. Li, W. Sun, L. Zhang, Y. Li, and Y. Rao, "Polarization-modulated random fiber laser," *Laser Phys. Lett.*, vol. 13, no. 5, pp. 055101 (2016).

[9] Y. Tang, and J. Xu, "A random Q-switched fiber laser," *Sci. Rep.*, vol. 5, pp. 9338 (2015).

[10] R. Ma, W. Zhang, X. Zeng, Z. Yang, Y. Rao, B. Yao, C. Yu, Y. Wu, and S. Yu, "Quasi mode-locking of coherent feedback random fiber laser," *Sci. Rep.*, vol. 6, pp. 39703 (2016).

[11] H. Zhang, H. Xiao, P. Zhou, X. Wang, and X. Xu, "Random distributed feedback Raman fiber laser with short cavity and its temporal properties," *IEEE Photon. Technol. Lett.*, vol. 26, no. 16, pp. 1605-1608 (2014).

[12] G. Ravet, A. A. Fotiadi, M. Blondel, and P. Mégret, "Passive Q-switching in all-fibre Raman laser with distributed Rayleigh feedback," *Electron. Lett.*, vol. 40, no. 9, pp. 528-529 (2004).

[13] G. P. Agrawal. Nonlinear Fiber Optics (Academic, 2013)

[14] A. K. Komarov, and K. P. Komarov, "Multistability and hysteresis phenomena in passive mode-locked lasers," *Phys. Review E.*, vol. 62, no. 6, pp. R7607 (2000).

[15] A. Zavyalov, R. Iliew, O. Eogrov, and F. Lederer, "Hysteresis of dissipative soliton molecules in mode-locked fiber lasers," *Opt. Lett.*, vol. 34, no. 24, pp. 3827-3829 (2009).